\documentstyle[12pt]{article}
\def\zid{1\kern-0.36em\llap~1}

\newcommand{\beq}{\begin{equation}}
\newcommand{\ber}{\begin{eqnarray}}
\newcommand{\eeq}{\end{equation}}
\newcommand{\eer}{\end{eqnarray}}

\begin{document}

\begin{titlepage}%\vbox {\vspace{0.1mm}} %Leaves space at top of 1st page.
\rightline{[SUNY BING 7/1/03 v2] }
 \rightline{ hep-ph/0308090}
\vspace{2mm}
%\vbox {\vspace{0.1mm}} %Leaves space at top of 1st page.
\vspace{2mm}
\begin{center}
{\bf \hspace{0.1 cm} PAIRING OF PARAFERMIONS OF ORDER 2: SENIORITY
MODEL }\\ \vspace{2mm} Charles A. Nelson\footnote{Electronic
address: cnelson @ binghamton.edu  }
\\ {\it Department of Physics, State University of New York at
Binghamton\\ Binghamton, N.Y. 13902}\\[2mm]
\end{center}

%\vspace{2mm}

\begin{abstract}

As generalizations of the fermion seniority model, four multi-mode
Hamiltonians are considered to investigate some of the
consequences of the pairing of parafermions of order two.
2-particle and 4-particle states are explicitly constructed for
$H_{A} \equiv -G\;A^{\dagger }A$ with $A^{\dagger } \equiv
\frac{1}{2} \sum_{m>0}\;c_{m}^{\dagger }c_{-m}^{\dagger }$ and the
distinct $H_{C}\equiv -G\;C^{\dagger }C$ with $C^{\dagger } \equiv
\frac{1}{2} \sum_{m>0}\;c_{-m}^{\dagger }c_{m}^{\dagger }$, and
for  the time-reversal invariant $H_{(-)}\equiv -G\;(A^{\dagger
}-C^{\dagger })(A-C)$ and $H_{(+)}\equiv -G\;(A^{\dagger
}+C^{\dagger })(A+C)$, which has no analogue in the fermion case.
The spectra and degeneracies are compared with those of the usual
fermion seniority model.

\end{abstract}

\end{titlepage}

\section{Introduction}

The physics of fermion pairing and fermion condensates [1] is
important in contemporary elementary particle physics in precision
QCD calculations for hadron spectroscopy (e.g. via lattice gauge
theory or chiral effective Lagrangians) and in research on
dynamical electro-weak symmetry breaking of the standard model
(e.g. via technicolor or a $ t \bar t$ condensate). In this paper,
we construct and study four Hamiltonians as generalizations of the
fermion seniority model [2] in order to investigate some of the
consequences associated with the pairing of parafermions of order
2. A physical significance of ``order $p = 2$" is that 2 or less
such parafermions can occur in the same quantum state. Usual
fermions correspond to $p=1$.

Although the idea of the possible existence of fundamental
particles associated with other representations of the permutation
group is an old and simple one [3], and despite the existence of
significant general results in relativistic local quantum field
theory concerning properties of elementary particles obeying
parastatistics [4,5,6], calculations in this field are sometimes
intractable because of algebraic complexities arising from the
basic tri-linear commutation relations, see (1-3), versus the
standard bi-linear commutation relations which occur in order
$p=1$. Order $p=2$ is indeed simpler than $p>2$ because there is a
``self-contained set" of 3 relations [4], see (7-9). With respect
to representations of the permutation group, consideration of the
``order $p = 2$" parafermions is not special in that there are
still the two $d=2$ dimensional representations, a mixed
representation still occurs at $d=3$, and multiple mixed
representations still occur at $d=4$.  On the other hand, for
$p>2$, mixed representation(s) with both totally-symmetric and
totally-anti-symmetric ones do occur, starting at $d=3$.

The first Hamiltonian considered in this paper is
$H_{A} \equiv -G\;A^{\dagger }A$ with $%
A \equiv \sum_{m>0}\;B^{(m)}$ where
$B^{(m)}=\frac{1}{2}c_{-m}c_{m}$. The mode index $k,l,m$ ranges
from $1$ to $\Omega $ and $\sum_{m>0}$ denotes summation over
$1,2,...\Omega $. In this paper, summation symbols are always
displayed, so repeated indices are not to be understood to be
summed. In many treatments of the usual $p=1$ seniority model, for
instance in nuclear physics applications, $\Omega $ is the number
of $(l,-l)$ pairs and the ``mode index" $k,l,m$ is the magnetic
quantum number. In this paper, the time-reversal operation, $T$,
will be analogously defined to exchange $\ l\;\leftrightarrow
\;-l$ , but except for the use of this exchange operation, no
explicit physical significance such as ``magnetic quantum number"
is associated with the $k,l,m$ index.

In paraquantization, it is instructive to begin by summarizing the
parafermi and parabose cases together:  The basic commuation
relations are
\begin{eqnarray}
\lbrack c_{k},[c_{l}^{\dagger },c_{m}]_{\mp }] &=&2\;\delta
_{kl}\;c_{m},  \\ \lbrack c_{k},[c_{l}^{\dagger },c_{m}^{\dagger
}]_{\mp }] &=&2\;\delta _{kl}\;c_{m}^{\dagger }\mp 2\;\delta
_{km}\;c_{l}^{\dagger },   \\ \lbrack c_{k},[c_{l},c_{m}]_{\mp }]
&=&0
\end{eqnarray}
following the standard convention that the upper (lower) signs
correspond respectively to \newline parafermions (parabosons). The
minus subscript is often suppressed so $[A,B]\equiv [A,B]_{-}
\equiv AB-BA$, and $\{A,B\}\equiv [A,B]_{+} \equiv AB+BA$. In this
paper, the corresponding creation and annihilation operators for
the
ordinary $p=1$ fermions are labeled $a_{k}^{\dagger }$ and $a_{l}$. For $%
p=1$, $a_{m}^{\dagger }a_{-m}^{\dagger }= - \, a_{-m}^{\dagger
}a_{m}^{\dagger }$ , but for $p>1$, $c_{m}^{\dagger
}c_{-m}^{\dagger }$ and $c_{-m}^{\dagger }c_{m}^{\dagger }$ are
distinct operators.\  The number operator for the para-particles
is $ N_{k}=\frac{1}{2}[c_{k}^{\dagger },c_{k}]_{\mp }\;\pm
\;\frac{p}{2} $ with the order of the paraparticles, $p$, being
the maximum number of parafermions (parabosons) in a totally
symmetric state (anti-symmetric state).  We assume a unique vacuum state with the usual properties $%
c_{k}|0>=0$, $<0|0>=1$, and $c_{k}c_{l}^{\dagger } |0>=p\;\delta
_{kl}|0>$.

From here on in this paper, the $c_{l}^{\dagger }$, and $c_{m}$
are parafermi operators of order 2. The following two commuting
operators frequently occur in this pairing analysis:
\begin{eqnarray} \widehat{N}\equiv \frac{1}{2}\sum_{m>0}\left( [c_{m}^{\dagger
},c_{m}]+[c_{-m}^{\dagger },c_{-m}]\right) \;+\;2\;\Omega \\%
\widehat{\rho }\equiv \frac{1}{2}\sum_{m>0}\left( \{c_{m}^{\dagger
},c_{m}\}-\{c_{-m}^{\dagger },c_{-m}\}\right)
\end{eqnarray}
Note that $\widehat{N}$ is the sum of the parafermion number
operators for the $2\Omega $ modes. However, although
$c_{l}^{\dagger }$, and $c_{m}$ are parafermi operators of order
2, $\widehat{\rho }$ has the formal structure of being the
difference of parabosonic number operators for the $m >0$ and
$m<0$ modes. The appearance of this T-odd operator $\widehat{\rho
}$ is a noteworthy difference versus the ordinary $p=1$ seniority
model, in which it vanishes.

Since $(B^{(m)})^{\dagger }=\frac{1}{2}c_{m}^{\dagger }
c_{-m}^{\dagger }$ and $ (D^{(m)})^{\dagger
}=\frac{1}{2}c_{-m}^{\dagger }c_{m}^{\dagger }$ are distinct
operators, we also consider a second Hamiltonian $H_{C}\equiv
-G\;C^{\dagger }C$ with $C \equiv \sum_{m>0}\;D^{(m)}$.  In $A$
and in $C$ the parafermions with $m$ and $-m$ are paired and so
states constructed as polynomials of $A$ and $C$ will be labeled
as states of seniority zero, $s=0$, since they are states built
out of paired particles. Thus, for $H_{A,C}$, the seniority $s$ is
the number of unpaired parafermions in the state, just as in the
$p=1$ case.

In the investigation and analysis of fermion pairing phenomena in
condensed matter physics, the presence or absence of time reversal
invariance and its consequences has been one of the important
symmetry considerations [7]. \ If, in this respect, one does treat
the $m$ index on $c_{m}$ as corresponding to a magnetic quantum
number, then under the time reversal operation, \ $
B^{(m)}\leftrightarrow $ $D^{(m)}$ and $H_{A\;}\leftrightarrow
H_{C}$ . Although there appears to be no obvious violation of
time-reversal invariance at the observable's level corresponding
to such a discrete switch of Hamiltonians, e.g. spectra will be
the same for the two respective Hamiltonians, we find this a
somewhat radical formal situation which appears not very easily
generalized, in particular with respect to inclusion of kinetic
energy terms and perturbations. Accordingly, in this paper we also
consider two time-reversal-invariant Hamiltonians which are mapped
into themselves by this time-reversal operation. \ These are
$H_{(-)}\equiv -G\;(A^{\dagger }-C^{\dagger })(A-C)$ and
$H_{(+)}\equiv -G\;(A^{\dagger }+C^{\dagger })(A+C)$. \ Note that
$H_{(+)}$ does not exist in the $p=1$ case. As will be discussed
below, results for $H_{(+)}$ such as its spectrum can normally be
obtained from those for $H_{(-)}$ by appropriate ``substitution
rules.''

The operators $\widehat{N}$ and $\widehat{\rho }$ commute with
$H_{A,C}$ but only  $\widehat{N}$, not $\widehat{\rho }$, commutes
with $H_{(\mp )}.$

As in the usual quasi-spin formalism [8], Section 2 of this paper
treats the algebras associated with the $H_{A,C}$  Hamiltonians,
$\widehat{N}$, $\widehat{\rho}$ ,$A$, $C$, and other such two-body
operators . Analogous to the $s=0$ operators $A$ and $C$, two sets
of $s=2$ two-body operators $B_{i}$ and $D_{i}$ with
$i=1,...,\Omega -1$ are introduced. In Sections 3 and 4, these
two-body operators $A, C, B_{i}$ and $D_{i}$  are used to
explicitly construct $N$-particle states with various senorities
($N$ is even). These results are used to study the spectrum for
$H_{A,C}$ and for $H_{\mp}$ in comparison with that of the fermion
seniority model $H\equiv -G\;\sum_{m>0}a_{m}^{\dagger
}a_{-m}^{\dagger }\sum_{l>0}a_{-l}a_{l}$ which has the $N-$state
spectrum
\begin{equation}
E_{s}(N)=-\frac{1}{4}G(N-s)(2\Omega _{p=1}-N-s+2);\quad
s=0,2,...N.
\end{equation}

The 2-particle and 4-particle states are explicitly constructed
for $H_{A,C}$ in Section 3 and for $H_{(-)}$ in Section 4. In both
cases for $s=4$ the construction of the 4-particle states is only
for $\Omega =4$.  For $H_{A,C}$ and for $H_{\mp}$, it is found
that the necessary mutual orthogonality properties of the
4-particle states are somewhat awkward to arrange using the
two-body operators $A$, $C$, $B_{i}$ and $D_{i}$.  There are two
built-in ``parafermi p-saturation'' conditions for $p=2$:
$(c_{k}^{\dagger })^{3}=0$ and $(A^{\dagger })^{M}=(C^{\dagger
})^{M}=0$ when $M=\Omega + 1 $ [this second fact is also true in
the $p=1$ case]. For some of the additional 4-particle states, for
instance, this has the consequence that some state normalization
constants vanish for small $\Omega$ values, because the states do
not then exist. Some results for arbitrary $\Omega$ beyond $N =4$
are derived. In Section 3, in all cases, the spectrum and
degeneracies for $H_{A,C}$ is found to be that of the usual $p=1$
senority model.

In Section 4, for $H_{(-)}$, for $N=2,4$ by explicit construction
of orthonormal states, a sizable number of additional states not
present in the analysis of $H_{A}$ are found to occur. For
$H_{(-)}$, results are obtained for arbitrary N-particles states
which can be constructed as polynomials in only the $A^{\dagger }$
and $\ C^{\dagger }$ operators. In all cases, the spectrum of
$H_{(-)}$ is found to be that of the $p=1$ senority model, except that $%
\Omega _{p=1}$ in (6) is replaced by $2\Omega $.  However, for
$H_{(-)}$ there are many additional degeneracies beyond those of
the usual $p=1$ senority model. These degeneracies can be
specified by an appropriate use of the seniority number, $s$.

A primary motivation for studying the seniority model is because
it is a simple model which has been used for fermions to
theoretically investigate and exhibit consequences of
fermion-pairing, of the microscopic realization of
superconductivity, and thereby of spontaneous symmetry breaking.
The most surprising result of this paper's analysis is that in
this multi-mode framework for parafermions of order two, it is
indeed possible to algebraically investigate the spectrum for each
of the four Hamiltonians. In hindsight, this tractability is
partially a consequence of three facts: (i) in a single mode, the
pairing-operators $B^{(m)} \equiv \frac{1}{2} c_{-m} c_{m} $ and
$D^{(m)}  \equiv \frac{1}{2} c_{m} c_{-m} $ separately lead to a
two-body operator, quasi-spin Hamiltonian structure which is
similar to that of the fermionic case, (ii) for different modes,
six Hermitian pairing-operators mutually commute: These operators
are ${ B_{1} }^{(m)} \equiv \frac{1}{2} ( {B^{(m)}}^{\dagger } +
{B^{(m)}} )$, ${ B_{2} }^{(m)} \equiv \frac{i}{2} (
{B^{(m)}}^{\dagger } - {B^{(m)}} )$, and ${ B_{3} }^{(m)} \equiv
\frac{1}{2} [ {B^{(m)}}^{\dagger }, {B^{(m)}} ]$ and analogously
for $ {D_{a} }^{(m)} $, and $
[B_{a}^{(l)},B_{b}^{(m)}]=[D_{a}^{(l)},D_{b}^{(m)}]=
[B_{a}^{(l)},D_{b}^{(m)}]=0$ for $l\neq m$ where $a,b=1,2,3$, and
(iii) for these operators in the same mode also $
[B_{a}^{(m)},D_{b}^{(m)}]=0$.

\section{Two-body operator algebras}

For parafermions of order 2, one has the relations
\begin{eqnarray}
c_{k}^{\dagger }c_{l}c_{m}+c_{m}c_{l}c_{k}^{\dagger } &=&2\;\delta
_{kl}\;c_{m} \\ c_{k}c_{l}^{\dagger }c_{m}+c_{m}c_{l}^{\dagger
}c_{k} &=&2\;\delta _{kl}\;c_{m}+2\;\delta _{lm}\;c_{k} \\
c_{k}c_{l}c_{m}+c_{m}c_{l}c_{k} &=&0
\end{eqnarray}
plus the Hermitian conjugate relations. Note from the left-hand
side of these relations that there is a simple
left$\leftrightarrow $right reordering symmetry. On the vacuum
state $c_{k}c_{l}^{\dagger } |0> =2  \delta _{kl} |0>$. A
consequence of (9) is the ``parafermi p-saturation'' conditions
noted in the ``Introduction".  Useful commutators involving
$c_{k}^{\dagger}$ , $ c_{-k}^{\dagger }$, $...$ pairs are in the
appendix of this paper.

It follows that the $s=0$ two-body operator $A\equiv
\sum_{m>0}\;B^{(m)}$ =$\frac{1}{2}\sum_{m>0}c_{-m}c_{m}$ has the
quasi-spin algebraic relations:
\begin{equation}
\lbrack A,A^{\dagger }]=-2Z_{A3},\quad \lbrack Z_{A3},A^{\dagger
}]=A^{\dagger },\quad \lbrack Z_{A3},A]=-A
\end{equation}
where $A^{\dagger }\equiv A_{1}+iA_{2}$ with $i=\sqrt {-1}$,
$A_{3}\equiv Z_{A3}$.   For $\overrightarrow{S_{A}}^{2}\equiv
(A_{1})^{2}+(A_{2})^{2}+(Z_{A3})^{2}$, one finds $[%
\overrightarrow{S_{A}}^{2},A_{1,2,3}]=0$ and $H_{A}=-G(\overrightarrow{S_{A}}%
^{2}-(Z_{A3})^{2}+Z_{A3})$.  Since $[A^{\dagger }A,Z_{A3}]=0$,
$H_{A}$ and $Z_{A3}$ can be simultaneously diagonalized. The $H_{A}$ eigenvalues are $E_{A}=-G%
\{s_{A}(s_{A}+1)-(z_{3A}$ )$^{2}+z_{3A}\}$.  The explicit
N-particle parafermion states of various seniorities corresponding
to this spectrum are constructed in Section 3 below. In terms
of the operators $\widehat{N}$ and $\widehat{%
\rho }$ defined by (4,5)
\begin{equation}
Z_{A3}\equiv \frac{1}{4}(\widehat{N}-\widehat{\rho }-2\Omega )
\end{equation}

Similarly, for
$C=\sum_{m>0}\;D^{(m)}=\frac{1}{2}\sum_{m>0}c_{m}c_{-m}$,
\begin{equation}
\lbrack C,C^{\dagger }]=-2Y_{C3},\quad \lbrack Y_{C3},C^{\dagger
}]=C^{\dagger }=C_{1}+iC_{2},\quad \lbrack Y_{C3},C]=-C
\end{equation}
where
\begin{equation}
Y_{C3}\equiv C_{3}\equiv \frac{1}{4}(\widehat{N}+\widehat{\rho
}-2\Omega ),
\end{equation}
Since $[C^{\dagger }C,Y_{C3}]=0$,  $H_{C}$ and $Y_{C3}$ can be
simultaneously diagonalized.  The $H_{C}$ eigenvalues are
$E_{A}=-G\{s_{C}(s_{C}+1)-(y_{3C}$ )$%
^{2}+y_{3C}\}$.

Useful eigenvalues for the vacuum state are
\begin{equation}
AA^{\dagger }|0>=CC^{\dagger }|0>=\Omega |0>,Z_{A3}|0>=Y_{C3}|0>=-\frac{1}{2}%
\Omega |0>
\end{equation}
and $\widehat{N}|0>=\widehat{\rho }|0>=0$. Unlike in the $p=1$
case, here for the $s=0$ N-particle states due to the occurrence of  $\widehat{\rho }$ as well as $\widehat{N}$, while $%
\widehat{N}(A^{\dagger })^{M}|0>=2M(A^{\dagger })^{M}|0>$ and $\widehat{N}%
(C^{\dagger })^{M}|0>=2M(C^{\dagger })^{M}|0>$ for $M$ a
non-negative integer, there is a minus sign in $\widehat{\rho
}(C^{\dagger })^{M}|0>=-2M(C^{\dagger })^{M}|0> $ versus
$\widehat{\rho }(A^{\dagger
})^{M}|0>=2M(A^{\dagger })^{M}|0>$. These states have respectively the energy eigenvalues $%
E_{0}^{(A,C)}(2M)=-GM(\Omega -M+1)$.

There are the useful commutators
\begin{eqnarray}
\lbrack \widehat{N},A^{\dagger }] &=&2A^{\dagger },\quad \lbrack \widehat{%
\rho },A^{\dagger }]=-2A^{\dagger } \\
\lbrack H_{A},A^{\dagger }] &=&-\frac{1}{2}G(\widehat{\rho }-\widehat{N}%
+4+2\Omega )A^{\dagger } \\ &=&-\frac{1}{2}GA^{\dagger
}(\widehat{\rho }-\widehat{N}+2\Omega )
\end{eqnarray}
The $\widehat{\rho }$ again appears with an extra minus sign in
the analogous expressions
\begin{eqnarray}
\lbrack \widehat{N},C^{\dagger }] &=&2C^{\dagger },\quad \lbrack \widehat{%
\rho },C^{\dagger }]=2C^{\dagger } \\
\lbrack H_{C},C^{\dagger }] &=&-\frac{1}{2}G(-\widehat{\rho }-\widehat{N}%
+4+2\Omega )C^{\dagger } \\ &=&-\frac{1}{2}GC^{\dagger
}(-\widehat{\rho }-\widehat{N}+2\Omega )
\end{eqnarray}
Note $[H_{A},C^{\dagger }]=[H_{C},A^{\dagger }]=0$.

To explicitly construct states with seniority $s\neq 0$, we define
the $s=2$ two-body operators
\begin{eqnarray}
B_{i}^{\dagger } &\equiv &\frac{1}{2}(\sum_{j=1}^{i}c_{j}^{\dagger
}c_{-j}^{\dagger }-ic_{(i+1)}^{\dagger }c_{-(i+1)}^{\dagger }) \\
&\equiv &\sum_{m=1}^{i}B^{(m)\dagger }-iB^{(i+1)\dagger };\quad
i=1,...,(\Omega -1)
\end{eqnarray}
and
\begin{equation}
D_{i}^{\dagger }\equiv \frac{1}{2}(\sum_{j=1}^{i}c_{-j}^{\dagger
}c_{j}^{\dagger }-ic_{-(i+1)}^{\dagger }c_{(i+1)}^{\dagger
});i=1,...,(\Omega -1)
\end{equation}
The 2-particle states $B_{i}^{\dagger }|0>$ and $D_{i}^{\dagger }|0>$ with $%
s=2$ respectively have zero $H_{A,C}$ energy eigenvalues,
$\widehat{N}$ eigenvalues of $2$, and respectively
$\widehat{\rho}$ eigenvalues of $\mp 2$ like respectively
$A^{\dagger }, C^{\dagger }$.  For $\Omega $ arbitrary,
\begin{equation}
\lbrack A,B_{i}^{\dagger }]=-2Z_{3Bi},\quad \lbrack A^{\dagger
},B_{i}^{\dagger }]=[B_{i}^{\dagger },B_{j}^{\dagger }]=0
\end{equation}
\begin{equation}
\lbrack H_{A},B_{i}^{\dagger }]=2GA^{\dagger }Z_{3Bi},\quad
\lbrack \widehat{N},B_{i}^{\dagger }]=2B_{i}^{\dagger },\quad
[\widehat{\rho},B_{i}^{\dagger }]= -2B_{i}^{\dagger }
\end{equation}
where
\begin{equation}
Z_{3Bi}\equiv \frac{1}{4}(\sum_{j=1}^{i}\{c_{j}^{\dagger
}c_{j}-c_{-j}c_{-j}^{\dagger }\}-i\{c_{(i+1)}^{\dagger
}c_{(i+1)}-c_{-(i+1)}c_{-(i+1)}^{\dagger }\})
\end{equation}
and
\begin{eqnarray}
\lbrack Z_{3Bi},A^{\dagger }] &=&B_{i}^{\dagger },\quad \lbrack
Z_{3A},B_{i}^{\dagger }]=B_{i}^{\dagger } \\ \lbrack
Z_{3A},Z_{3Bi}] &=&[Z_{3Bi},Z_{3Bj}]=0
\end{eqnarray}
On the vacuum state, $Z_{3Bi}|0>=0$, so $AB_{i}^{\dagger
}|0>=B_{i}A^{\dagger }|0>=0$. Alternatively, in terms of mode
operators
\begin{eqnarray}
Z_{3Bi} &\equiv &\frac{1}{4}(\sum_{m=1}^{i}\{\widehat{N}_{B}^{(m)}-\widehat{%
\rho }_{B}^{(m)}\}-i\{\widehat{N}_{B}^{(i+1)}-\widehat{\rho
}_{B}^{(i+1)}\})
\\
&=&\sum_{m=1}^{i}Z_{3}^{(m)}-iZ_{3}^{(i+1)}
\end{eqnarray}
where
\begin{eqnarray}
\widehat{N}_{B}^{(m)} &\equiv &\frac{1}{2}\left( [c_{m}^{\dagger
},c_{m}]+[c_{-m}^{\dagger },c_{-m}]\right) \;+\;2\;\Omega  \\
\widehat{\rho }_{B}^{(m)} &\equiv &\frac{1}{2}\left(
\{c_{m}^{\dagger },c_{m}\}-\{c_{-m}^{\dagger },c_{-m}\}\right) \;
\\ Z_{3}^{(m)} &\equiv &\frac{1}{4}\left( c_{m}^{\dagger
}c_{m}-c_{-m}c_{-m}^{\dagger }\right)
\end{eqnarray}

For $\Omega > 2$, these operators do not completely close at the
level of the $A$'s,$C$'s,$B_{i}$'s,$D_{i}$'s but instead involve
mode operators:
\begin{equation}
\lbrack Z_{3Bi},B_{j}^{\dagger }]=\left\{
\begin{array}{c}
\sum_{m=1}^{i}B^{(m)\dagger }+(i)^{2}B^{(i+1)\dagger },i=j \\
B_{i<}^{\dagger },\quad i\neq j
\end{array}
\right\}
\end{equation}
where $i<$ denotes the smaller of $i,j$.

Similarly,
\begin{equation} \lbrack C,D_{i}^{\dagger
}]=-2Y_{3Di},\quad \lbrack C^{\dagger },D_{i}^{\dagger
}]=[D_{i}^{\dagger },D_{j}^{\dagger }]=0
\end{equation}
\begin{equation}
\lbrack H_{C},D_{i}^{\dagger }]=2GC^{\dagger }Y_{3Di},\quad
\lbrack \widehat{N},D_{i}^{\dagger }]=2D_{i}^{\dagger },\quad
[\widehat{\rho},D_{i}^{\dagger }]= 2D_{i}^{\dagger }
\end{equation}
where
\begin{eqnarray}
Y_{3Di} &\equiv &\frac{1}{4}(\sum_{j=1}^{i}\{c_{-j}^{\dagger
}c_{-j}-c_{j}c_{j}^{\dagger }\}-i\{c_{-(i+1)}^{\dagger
}c_{-(i+1)}-c_{(i+1)}c_{(i+1)}^{\dagger }\}) \\
&=&\sum_{m=1}^{i}Y_{3}^{(m)}-iY_{3}^{(i+1)}
\end{eqnarray}
and
\begin{eqnarray}
\lbrack Y_{3Di},C^{\dagger }] &=&D_{i}^{\dagger },\quad \lbrack
Y_{3C},D_{i}^{\dagger }]=D_{i}^{\dagger } \\ \lbrack
Y_{3C},Y_{3Di}] &=&[Y_{3Di},Y_{3Dj}]=0
\end{eqnarray}
\begin{equation}
\lbrack Y_{3Di},D_{j}^{\dagger }]=\left\{
\begin{array}{c}
\sum_{m=1}^{i}D^{(m)\dagger }+(i)^{2}D^{(i+1)\dagger },i=j \\
D_{i<}^{\dagger },\quad i\neq j
\end{array}
\right\}
\end{equation}
On the vacuum state $Y_{3Di}|0>=0$, so $CD_{i}^{\dagger
}|0>=D_{i}C^{\dagger }|0>=0$, and
\begin{eqnarray}
Y_{3Di} &\equiv &\frac{1}{4}(\sum_{m=1}^{i}\{\widehat{N}_{B}^{(m)}+\widehat{%
\rho }_{B}^{(m)}\}-i\{\widehat{N}_{B}^{(i+1)}+\widehat{\rho
}_{B}^{(i+1)}\})
\\
&=&\sum_{m=1}^{i}Y_{3}^{(m)}-iY_{3}^{(i+1)}
\end{eqnarray}
where $Y_{3}^{(m)}\equiv \frac{1}{4}\left( c_{-m}^{\dagger
}c_{-m}-c_{m}c_{m}^{\dagger }\right) .$ The states $B_{i}^{\dagger
}|0>,A^{\dagger }|0>$ have $Z_{A3}$ eigenvalues of
$(1-\frac{1}{2}\Omega )$, and $Y_{C3}$ eigenvalues of
$(-\frac{1}{2}\Omega )$; similarly $D_{i}^{\dagger }|0>,C^{\dagger
}|0>$ have $Y_{C3}$ eigenvalues of $(1-\frac{1}{2}\Omega )$, and
$Z_{A3}$ eigenvalues of $(-\frac{1}{2}\Omega )$.

Note that
\begin{equation}
\lbrack B_{i},B_{j}^{\dagger }]=-2Z_{3Bi<}-\frac{1}{2}\delta _{ij}
i(i+1)\{c_{(i+1)}^{\dagger
}c_{(i+1)}-c_{-(i+1)}c_{-(i+1)}^{\dagger }\}
\end{equation}
where again $i<$ denotes the smaller of $i,j$; the last term's
factor also appears in (26). On the vacuum state
$B_{i}B_{j}^{\dagger }|0>=i(i+1)\delta _{ij}||0>$, so
$B_{i}^{\dagger }|0>$ are orthogonal for different $i$ values.
Similarly,
\begin{equation}
\lbrack D_{i},D_{j}^{\dagger }]=-2Y_{3Di<}-\frac{1}{2}\delta _{ij}
i(i+1)\{c_{-(i+1)}^{\dagger
}c_{-(i+1)}-c_{(i+1)}c_{(i+1)}^{\dagger }\}
\end{equation}
and $D_{i}D_{j}^{\dagger }|0>=i(i+1)\delta _{ij}||0>$.

\section{Spectrum of $H_A$}

Useful relations for treating arbitrary N particle states include: for $%
r=1,2,...$
\begin{eqnarray}
A(A^{\dagger })^{r}|0 &>&=r(\Omega -r+1)(A^{\dagger })^{r-1}|0> \\
(A)^{r}(A^{\dagger })^{r}|0 &>&=r!\Omega (\Omega -1)\cdots (\Omega
-r+1)|0>
\\
Z_{A3}(A^{\dagger })^{r}|0 &>&=(r-\frac{1}{2}\Omega )(A^{\dagger
})^{r}|0>
\end{eqnarray}
Since we find the $H_A$ case to be relatively simple, for instance
it has the same spectrum as in the usual $p=1$ senority model, we
do not evaluate normalization constants in this section.

For arbitrary $N\geq 2$, the $s=0$ states $|N^A>_{0}=(A^{\dagger })^{N/2}|0>$ have energy eigenvalues $E_{0}^{(A)}(N)=-\frac{1}{4%
}GN(2\Omega -N+2)$.

For the $s=2$, N-particle states with $N\geq 2$, $|N^A>_{2}\equiv
B_{i}^{\dagger }(A^{\dagger })^{(N-2)/2}|0>$ have $E_{2}^{(A)}(N)=-\frac{1}{4}%
G(N-2)(2\Omega -N)$.

For $s=4$, we are interested in both the $H_{A}$ states for
themselves and also for comparison below with those occurring in
the analysis of the $H_{(-)}$ states:  For $\Omega =4 $, the
explicit orthogonal 4-particle states include two with zero
$H_{A}$ eigenvalues. From the analogous $p=1
$ spectra and using completeness, we classify them as $s=4$ states. These two states are $|4_{a}^{A}>_{4}=\frac{2}{3}%
B_{1}^{\dagger }(B_{3}^{\dagger }-B_{2}^{\dagger })|0>$ , $|4_{b}^{A}>_{4}=%
\frac{1}{\sqrt{3}}B_{2}^{\dagger }(B_{3}^{\dagger }+A^{\dagger
})|0>$; to achieve orthogonality these states are somewhat
complicated in terms of the $B_{i}^{\dagger }$ operators. The
other $\Omega =4$ states are special cases of ones discussed
above: $|4^{A}>_{0}=(A^{\dagger })^{2}|0>$ ,
$|4_{i}^{A}>_{2}=A^{\dagger }B_{i}^{\dagger }|0>$ where $i=1,2,3
$.

Next, for the $s=4$, N-particle states with $N\geq 4$, we consider
\newline $|N^A>_{4}\equiv B_{i}^{\dagger }B_{j}^{\dagger
}A^{\dagger })^{(N-4)/2}|0>$ with $i<j$. \ Note that it  is at the
$s=4$ senority level that in using the $B_{i}^{\dagger }$
operators, the orthogonality requirements started to produce
complications for the $H_{A}$ case, and similarly in the $H_{(-)}$
case below. So in considering only $|N^A>_{4}$, we are ignoring
this difficulty. When $H_{A}$ operates on this state, one can
commute $A$ past the first $B_{i}^{\dagger }$ using
$[A,B_{i}^{\dagger }]=-2Z_{3Bi}$. In commuting $Z_{3Bi}$ past the
next $B_{j}^{\dagger }$ one can use (34), $\ \
[Z_{3Bi},B_{j}^{\dagger }]=B_{i}^{\dagger }$ $\ \sin $ce \ $i<j$,
\ however, $B_{i}^{\dagger }$  produces an $(N-2)$-particle state.
\ We proceed by dropping such terms because such $(N-2)$ terms
will not
contribute if the $H_{A}$ expectation value is calculated. In this manner, we obtain $%
E_{4}^{(A)}(N)=-\frac{1}{4}G(N-4)(2\Omega -N-2).$  This is not a
complete derivation because the $|N^A>_{4}$ are not mutually
orthogonal. If we proceed similarly, for arbitrary senority
$s=2t$, then for N-particle states with $N\geq s$,
$|N^A>_{s}\equiv \{B_{i_{1}}^{\dagger }B_{1_{2}}^{\dagger }\cdots
B_{i_{t}}^{\dagger }\}(A^{\dagger })^{(N-s)/2}|0>
$ with $i_{1}<i_{2}<\cdots <i_{t}$, we obtain  $%
E_{s}^{(A)}(N)=-\frac{1}{4}G(N-s)(2\Omega -N-s+2);\quad s=0,2,...N
$  by induction.

\section{Analysis of $H_{(-)}$}

For the Hamiltonians $H_{(\mp )}$ whereas $[\widehat{N},H_{(\mp
)}]=0$ ,
\begin{equation}
\lbrack \widehat{\rho },H_{(\mp )}]=\mp 4G(A^{\dagger
}C-C^{\dagger }A)=\mp 4(H_{AC}-H_{CA})
\end{equation}
so the N-particle eigenstates of  $H_{(\mp )}$ can no longer be
classified by the eigenvalues of $\widehat{\rho }.$  Note that
$H_{(\mp )}=H_{A}\pm H_{AC}\pm H_{CA}+H_{C}$ where
$H_{AC}=GA^{\dagger }C^{\dagger }$ and $H_{CA}=GC^{\dagger
}A^{\dagger }$.

The algebra associated with $H_{(-)}$ includes the equations
\begin{eqnarray}
\lbrack A\pm C,A^{\dagger }\pm C^{\dagger }] &=&-N_{AC3},\quad
[A+C,A^{\dagger }-C^{\dagger }]=\widehat{\rho }
\\ \lbrack \widehat{N},A^{\dagger }\mp C^{\dagger }]
&=&2(A^{\dagger }\mp C^{\dagger }),\quad [\widehat{\rho
},A^{\dagger }\mp C^{\dagger }]=-2(A^{\dagger }\pm C^{\dagger })
\end{eqnarray}
where $N_{AC3}=(\widehat{N}-2\Omega )=2Z_{A3}+2Y_{C3}=\frac{1}{2}%
\sum_{m>0}\left( [c_{m}^{\dagger },c_{m}]+[c_{-m}^{\dagger
},c_{-m}]\right) $ includes the zero point energy.  In comparison,
note that $\widehat{\rho }=- 2Z_{A3}+2Y_{C3}$. Also
\begin{eqnarray}
\lbrack H_{(-)},A^{\dagger }-C^{\dagger }] &=&G(A^{\dagger
}-C^{\dagger })N_{AC3} \\
\lbrack H_{(-)},A^{\dagger }+C^{\dagger }] &=&-G(A^{\dagger }-C^{\dagger })%
\widehat{\rho }
\end{eqnarray}

\indent {\bf $N =2, 4 $ particle states:}

We list the orthogonal 2-particle states, their energy
eigenvalues, and
normalization constants $\mathit{N}_{s}\mathit{(}N\mathit{)\equiv }$ $%
_{s}<N|N>_{s}$:

Ones with $s=0,$

$|2^{-}>_{0}=(A^{\dagger }-C^{\dagger })|0>$, $E_{0}^{(-)}(2)=-2G\Omega $; $%
{N}_{0}{(2)}=2\Omega $

$|\breve{2}^{-}>_{0}=(A^{\dagger }+C^{\dagger })|0>$, $E_{0}^{(-)}(\breve{2}%
)=0$; ${N}_{0}{(\breve{2})}=2\Omega $

Ones with $s=2,$

$|2_{i}^{-}>_{2}=(B_{i}^{\dagger }-D_{i}^{\dagger })|0>$, $E_{2}^{(-)}(2)=0$%
; ${N}_{2}{(2)}=2i(i+1)$

$|\breve{2}_{i}^{-}>_{2}=(B_{i}^{\dagger }+D_{i}^{\dagger })|0>$, $%
E_{2}^{(-)}(\breve{2})=0$; ${N}_{2}{(\breve{2})}=2i(i+1)$

In calculating degeneracies, it is to be understood that
$i,j=1,2,...,\Omega -1$. \

The additional states not present in the $H_{A,C}$ cases are
denoted with a ``breve'', or ``short vowel'', accent. Such states
are therefore not analogous to ones in the usual $p=1$ seniority
model.

We similarly list the orthogonal 4-particle states:

Ones with $s=0,$

$|4^{-}>_{0}=(A^{\dagger }-C^{\dagger })^{2}|0>$, $E_{0}^{(-)}(4)=-2G(2%
\Omega -1)$; ${N}_{0}{(4)}=2\Omega (4\Omega -1)$

$|\breve{4}^{-}>_{0}=(A^{\dagger }-C^{\dagger })(A^{\dagger
}+C^{\dagger
})|0>$, $E_{0}^{(-)}(\breve{4})=-2G(\Omega -1)$; ${N}_{0}{(%
\breve{4})}=2(\Omega )^{2}$

$|\breve{4}_{1}^{-}>_{0}=(\{A^{\dagger }\}^{2}+\{C^{\dagger }\}^{2}+2(\frac{%
\Omega -1}{\Omega })A^{\dagger }C^{\dagger })|0>$, $E_{0}^{(-)}(\breve{4}%
_{1})=0$; ${N}_{0}{(\breve{4}}_{1}{)}=4(2\Omega -1)(\Omega -1)$

For instance in $|\breve{4}_{1}^{-}>_{0}$, a ``number subscript''
is used on the additional states label to denote ones constructed
with an $\Omega $ dependence so as to achieve orthogonality. \ Due
to the parafermi saturation such a state is absent for $\Omega $
sufficiently small; this is seen in the
norm vanishing and in the vanishing of a factor like $(\frac{\Omega -1}{%
\Omega })$. \ Completeness in each ``$N,s$ sector'' is achieved by
construction.

Ones with $s=2,$ which are orthogonal for $i\neq j$ ,

$|4_{i}^{-}>_{2}=(B_{i}^{\dagger }-D_{i}^{\dagger })(A^{\dagger
}-C^{\dagger
})|0>$, $E_{2}^{(-)}(4)=-2G(2\Omega -2)$; ${N}_{2}{(4)}%
=4(\Omega -1)i(i+1)$

$|\breve{4}_{i}^{-}>_{2}=(B_{i}^{\dagger }+D_{i}^{\dagger
})(A^{\dagger
}-C^{\dagger })|0>$, $E_{2}^{(-)}(\breve{4})=-2G(2\Omega -2)$; ${N}%
_{2}{(\breve{4})}=4(\Omega -1)i(i+1)$

$|\breve{4}_{i1}^{-}>_{2}=(\Omega \{B_{i}^{\dagger }A^{\dagger
}+D_{i}^{\dagger }C^{\dagger }\}+(\Omega -2)\{B_{i}^{\dagger
}C^{\dagger
}+D_{i}^{\dagger }A^{\dagger }\})|0>$, $E_{2}^{(-)}(\breve{4}_{1})=0$; $%
{N}_{2}{(\breve{4}}_{i1}{)}=4\Omega (\Omega -1)(\Omega -2)i(i+1)$

$|\breve{4}_{i2}^{-}>_{2}=(\Omega \{B_{i}^{\dagger }A^{\dagger
}-D_{i}^{\dagger }C^{\dagger }\}+(\Omega -2)\{B_{i}^{\dagger
}C^{\dagger
}-D_{i}^{\dagger }A^{\dagger }\})|0>$, $E_{2}^{(-)}(\breve{4}_{2})=0$; $%
{N}_{2}{(\breve{4}}_{i2}{)}=4\Omega (\Omega -1)(\Omega -2)i(i+1)$

Ones with $s=4,$ for the case $\Omega =4$; all with zero energy
eigenvalues and the same normalization constant
${N}_{4}{(4)}=144$,

$|4^{-}>_{4}=(B_{1}^{\dagger }-D_{1}^{\dagger })(B_{3}^{\dagger
}-B_{2}^{\dagger }-\{D_{3}^{\dagger }-D_{2}^{\dagger }\})|0>$;

$|\breve{4}_{a}^{-}>_{4}=(B_{1}^{\dagger }-D_{1}^{\dagger
})(B_{3}^{\dagger }-B_{2}^{\dagger }+\{D_{3}^{\dagger
}-D_{2}^{\dagger }\})|0>;$

$|\breve{4}_{b}^{-}>_{4}=(B_{1}^{\dagger }+D_{1}^{\dagger
})(B_{3}^{\dagger }-B_{2}^{\dagger }-\{D_{3}^{\dagger
}-D_{2}^{\dagger }\})|0>;$

$|\breve{4}_{c}^{-}>_{4}=(B_{1}^{\dagger }+D_{1}^{\dagger
})(B_{3}^{\dagger }-B_{2}^{\dagger }+\{D_{3}^{\dagger
}-D_{2}^{\dagger }\})|0>;$

Plus two analogues of $H_{A}$ states, \ and two additional states,

$|4_{\alpha }^{-}>_{4}=(B_{2}^{\dagger }\{B_{3}^{\dagger
}+A^{\dagger }\}+D_{2}^{\dagger }\{D_{3}^{\dagger }+C^{\dagger
}\})|0>$; ${N}_{4}{(4}_{\alpha }{)}=336$

$|\breve{4}_{\alpha }^{-}>_{4}=(B_{2}^{\dagger }\{B_{3}^{\dagger
}+A^{\dagger }\}-D_{2}^{\dagger }\{D_{3}^{\dagger }+C^{\dagger
}\})|0>$; ${N}_{4}{(\breve{4}}_{\alpha }{)}=336$

$|4_{\beta }^{-}>_{4}=\frac{1}{\sqrt{2}}(B_{2}^{\dagger
}D_{3}^{\dagger }+D_{2}^{\dagger }B_{3}^{\dagger })|0>$;
${N}_{4}{(4}_{\beta }{)}=144$

$|\breve{4}_{\beta }^{-}>_{4}=\frac{1}{\sqrt{2}}(B_{2}^{\dagger
}D_{3}^{\dagger }-D_{2}^{\dagger }B_{3}^{\dagger })|0>$; ${N}_{4}{(\breve{4}}%
_{\beta }{)}=144$

\indent {\bf Some $N$-particle states constructed as polynomials
in $A^{\dagger }$ and $C^{\dagger }$ :}

Useful relations for treating arbitrary N particle states include: for $%
r=1,2,...$
\begin{eqnarray}
(A-C)(A^{\dagger }-C^{\dagger })^{r}|0 &>&=r(2\Omega
-r+1)(A^{\dagger }-C^{\dagger })^{r-1}|0>
\end{eqnarray}
$$ \displaylines{
 (A\mp C)^{r}(A^{\dagger }\mp C^{\dagger })^{r}|0>=r!
\sum_{t=0}^{r} { r \choose t} \Omega (\Omega -1)\cdots (\Omega
-r+t+1) \hfill \qquad\cr \Omega (\Omega -1)\cdots (\Omega -t+1)
 |0>\qquad\cr } $$
 and
\begin{eqnarray}
\widehat{N} (A^{\dagger })^{r_1} (C^{\dagger })^{r_2}|0> =2( r_1 +
r_2) (A^{\dagger })^{r_1} (C^{\dagger })^{r_2}|0> \\
\widehat{\rho} (A^{\dagger })^{r_1} (C^{\dagger })^{r_2}|0> = - 2(
r_1 - r_2) (A^{\dagger })^{r_1} (C^{\dagger })^{r_2}|0>
\end{eqnarray}

Ignoring the mutual orthogonality requirement, for arbitrary N,
the following seniority $s_{p=1}=0$ states are found to have the
following associated eigenvalues:

For $N\geq 2$, $(A^{\dagger }-C^{\dagger })^{N/2}|0>$ has $%
E_{0}^{(-)}(N)=-\frac{1}{4}GN(4\Omega -N+2)$.

For $\breve{N}\geq 2$, $(A^{\dagger }-C^{\dagger
})^{(N-2)/2}(A^{\dagger }+C^{\dagger })|0>$ has
$E_{0}^{(-)}(\breve{N})=-\frac{1}{4}G(N-2)(4\Omega -N)$.

For $\breve{N}^{^{\prime }}\geq 4$,  $%
(A^{\dagger }-C^{\dagger })(A^{\dagger }+C^{\dagger
})^{(N-2)/2}|0>$ also has $ E_{0}^{(-)}(\breve{N}^{^{\prime
}})=-\frac{1}{4}G(N-2)(4\Omega -N)$. So for $N\geq 4$, both these
states are degenerate with the $s=2$ states with N-particles.

For $\breve{N}^{^{^{\prime \prime }}}\geq 4$, $(A^{\dagger
}-C^{\dagger
})^{(N-4)/2}(\{A^{\dagger }\}^{2}+\{C^{\dagger }\}^{2}+2\{\frac{\Omega -1}{%
\Omega }\}A^{\dagger }C^{\dagger })|0>$  has $E_{0}^{(-)}(\breve{N}%
^{^{\prime \prime }})=-\frac{1}{4}G(N-4)(4\Omega -N-2)$ so this
state is degenerate with the $s=4$ state with N-particles.

For these states, ignoring orthogonality, the $H_{(-)}$ spectrum
is found to be the same as that of $H_{A}$, except that $ \Omega
_{p=1}$ in (6) is replaced by $2\Omega $. There are, however, many
additional degeneracies which can be counted via completeness in
each the $N,s_{p=1}$ sector, for $\Omega$ sufficiently large so
that the absence of states due to p-saturation can be ignored.

The states of the $H_{(+)}$ spectrum follow isomorphically by
letting $C \rightarrow - C$ and $D_{i}^{\dagger } \rightarrow -
D_{i}^{\dagger }$:  for instance,  $|2^{+}>_{0}=(A^{\dagger
}+C^{\dagger })|0>$ has $E_{0}^{(+)}(2)=-2G\Omega $;
$|\breve{2}^{+}>_{0}=(A^{\dagger }-C^{\dagger })|0>$ has
$E_{0}^{(+)}(\breve{2} )=0$; $|2_{i}^{+}>_{2}=(B_{i}^{\dagger
}+D_{i}^{\dagger })|0>$ has $E_{2}^{(+)}(2)=0$ ; and
$|\breve{2}_{i}^{+}>_{2}=(B_{i}^{\dagger }-D_{i}^{\dagger })|0>$
has $ E_{2}^{(+)}(\breve{2})=0$.

{\bf Acknowledgments: }

We thank Terence Tarnowsky for his participation in the initial
part of this research, and thank the Fermilab Theory Group for an
intellectually stimulating visit during the summer of 2002. This
work was partially supported by U.S. Dept. of Energy Contract No.
DE-FG 02-86ER40291.

{\bf Appendix:  ${c_k}^{\dagger}$, ${c_{-k}}^{\dagger }$,$...$
pair commutators }

Although equations (7-9) are sufficient, the following commutators
are algebraically sometimes more direct or useful as checks:
\begin{eqnarray}
\lbrack c_{k}c_{-k},c_{l}^{\dagger }c_{-l}^{\dagger }]
&=&2\;\delta _{-k,l}(c_{k}c_{k}^{\dagger }-c_{-k}^{\dagger
}\;c_{-k}) \\ \lbrack c_{k}^{\dagger }c_{-k}^{\dagger
},c_{l}^{\dagger }c_{-l}^{\dagger }] &=&0
\end{eqnarray}
From the latter it follows that $\
[B^{(m)},B^{(l)}]=[D^{(m)},D^{(l)}]=[B^{(m)},D^{(l)}]=0$, so  $%
[A,C]=[B_{i},B_{j}]=[D_{i},D_{j}]=[B_{i},D_{j}]=0$.
\begin{eqnarray}
\lbrack c_{k}^{\dagger }c_{k},c_{l}^{\dagger }c_{-l}^{\dagger }]
&=&2\;\delta _{k,l}c_{k}^{\dagger }\;c_{-k}, \quad
[c_{k}c_{k}^{\dagger },c_{l}^{\dagger }c_{-l}^{\dagger
}]=-2\;\delta _{k,-l}c_{-k}^{\dagger }\;c_{k} \\ \lbrack
c_{k}^{\dagger }c_{k},c_{l}^{\dagger }c_{l}] &=&[c_{k}^{\dagger
}c_{k},c_{l}c_{l}^{\dagger }]=0
\end{eqnarray}
The mode operators $B^{(m)}$ also satisfy $[B^{(m)},B^{(m)\dagger
}]=-2Z_{3}^{(m)},\quad \lbrack Z_{3}^{(m)},B^{(m)\dagger
}]=B^{(m)\dagger }$, where $B^{(m)\dagger }\equiv
B_{1}^{(m)}+iB_{2}^{(m)}$ and $B_{3}^{(m)}\equiv Z_{3}^{(m)}$. See
also (31-33). Also, $[D^{(m)},D^{(m)\dagger }]=-2Y_{3}^{(m)},
\newline  \lbrack Y_{3}^{(m)},D^{(m)\dagger }]=D^{(m)\dagger }$.
Thus, for both the $B^{(m)}$'s and $D^{(m)}$'s one has a two-body
operator, quasi-spin, and single-mode Hamiltonian structure,
analogous to those at the $A$'s and $C$'s level.

Some of the commutators vanish at the $A$'s and $C$'s level
because $\lbrack B_{a}^{(l)},B_{b}^{(m)}]|_{l\neq m}= \newline
[D_{a}^{(l)},D_{b}^{(m)}]|_{l\neq m}=[B_{a}^{(l)},D_{b}^{(m)}]=0$
for $a,b=1,2,3$. Thus, $[A,C^{\dagger }]=[B_{i},D_{i}^{\dagger
}]=[A,D_{i}^{\dagger }]=[A,D_{i}]=[C,B_{i}^{\dagger}]=[C,B_{i}]=0
$, and $ [Z_{A3},Y_{C3}]=[Z_{3Bi},Y_{C3}]=[Z_{A3},Y_{3Di}]
=[Z_{3Bi},Y_{3Di}]=0$. Also, $\ [Z_{A3},C^{\dagger
}]=[Z_{3Bi},C^{\dagger }]=[Z_{A3},D_{i}^{\dagger
}]=[Z_{3Bi},D_{i}^{\dagger }]=0$, $ [Y_{C3},A^{\dagger
}]=[Y_{3Di},A^{\dagger }]=[Y_{C3},B_{i}^{\dagger
}]=[Y_{3Di},B_{i}^{\dagger }]=0$ , and their adjoints vanish.

\end{document}